\documentclass[fleqn,10pt]{wlscirep}
\usepackage[utf8]{inputenc}
\usepackage[T1]{fontenc}
\usepackage{graphicx}%
\usepackage{multirow}%
\usepackage{amsmath,amssymb,amsfonts}%
\usepackage{amsthm}%
\usepackage{mathrsfs}%
\usepackage[title]{appendix}%
\usepackage{xcolor}%
\usepackage{textcomp}%
\usepackage{manyfoot}%
\usepackage{booktabs}%
\usepackage{algorithm}%
\usepackage{algorithmicx}%
\usepackage{algpseudocode}%
\usepackage{listings}%
\usepackage{booktabs,caption}
\usepackage{subcaption}
\usepackage{lineno}
\usepackage{multirow} 
\usepackage{rotating} 
\usepackage{adjustbox}
\usepackage{siunitx}
\newcolumntype{Y}{>{\centering\arraybackslash}X}

\begin{document}

\title{Actualised and future changes in regional economic growth through sea level rise}

\author[1,2*]{Theodoros Chatzivasileiadis}
\author[1]{Ignasi Cortés Arbués}
\author[3]{Jochen Hinkel}
\author[3]{Daniel Lincke}
\author[4,5,6,7,8,9,10]{Richard S.J. Tol}
\affil[1]{Department of Multi-Actor Systems, Faculty of Technology, Policy and Management, Delft University of Technology, Delft, The Netherlands}
\affil[2]{PBL\textemdash Netherlands Environmental Assessment Agency, The Hague, The Netherlands}
\affil[3]{GCF\textemdash Global Climate Forum, Berlin, Germany}
\affil[4]{Department of Economics, University of Sussex, Falmer, United Kingdom}
\affil[5]{Institute for Environmental Studies, Vrije Universiteit, Amsterdam, The Netherlands}
\affil[6]{Department of Spatial Economics, Vrije Universiteit, Amsterdam, The Netherlands}
\affil[7]{Tinbergen Institute, Amsterdam, The Netherlands}
\affil[8]{CESifo, Munich, Germany}
\affil[9]{Payne Institute for Public Policy, Colorado School of Mines, Golden, CO, USA}
\affil[10]{College of Business, Abu Dhabi University, UAE}
\affil[*]{Corresponding author: t.chatzivasileiadis@tudelft.nl}

\begin{abstract}
This study investigates the long-term economic impact of sea-level rise (SLR) on coastal regions in Europe, focusing on Gross Domestic Product (GDP). Using a novel dataset covering regional SLR and economic growth from 1900 to 2020, we quantify the relationships between SLR and regional GDP per capita across 79 coastal EU\&UK regions. Our results reveal that the current SLR has already negatively influenced GDP of coastal regions, leading to a cumulative 4.7\% loss at 39 cm of SLR. Over the 120 year period studied, the actualised impact of SLR on the \emph{annual} growth rate is between -0.02\% and 0.04\%. Extrapolating these findings to future climate and socio-economic scenarios, we show that in absence of additional adaptation measures, GDP losses by 2100 could range between -6.3\% and -20.8\% under the most extreme SLR scenario (SSP5-RCP8.5 High-end Ice, or -4.0\% to -14.1\% in SSP5-RCP8.5 High Ice). This statistical analysis utilising a century-long dataset, provides an empirical foundation for designing region-specific climate adaptation strategies to mitigate economic damages caused by SLR. Our evidence supports the argument for strategically relocating assets and establishing coastal setback zones when it is economically preferable and socially agreeable, given that protection investments have an economic impact.

\end{abstract}


\maketitle
\newpage

\section*{Introduction}
The escalating threats of climate change pose a profound risk to the global economy. These adverse impacts of climate change are notably asymmetric across regions. The concern is particularly pressing in the context of sea-level rise (SLR), which unevenly affects areas where significant populations and productive capital are concentrated: coastal cities and regions \cite{mcgranahan2007rising,smith2011we}. These regions experience rapid population growth and contribute substantially to Gross Domestic Product (GDP)\cite{copernicus2023}, despite being at the high risk of coastal flooding and substantial economic damage \cite{bosello2012,parrado2020fiscal}. 

Recent research has identified coastal impacts from SLR as one of the principal economic costs imposed by climate change (e.g., \cite{tol1, vousdoukas2018climatic}), using predominately ex-ante analyses. These impacts are second only to health effects and exceed the impacts on agriculture, energy sectors, and riverine flooding \cite{ciscar2014climate, schinko2020economy}. To keep coastal flood damages at acceptable levels compared to the overall economy, it is imperative to enhance flood defence systems that can withstand a rise in sea levels between 0.5 and 2.5 metres \cite{vousdoukas2018climatic} or to consider transformational adaptation, like planned relocation of people and economic activities in some regions \cite{haasnoot2021pathways}. Under extreme SLR, an area covering 0.5 to 0.7\% of global land - in many cases, the most economically productive land - will be exposed to occasional coastal flooding from a 1 in 100-year return period event by the year 2100 \cite{kirezci2020projections}. This could affect 2.5\% to 4.1\% of the world's population and damage assets equivalent to 12\% to 20\% of global GDP. Similarly, in the absence of incremental or transformational adaptation measures, anticipated \textit{direct} annual losses stemming from coastal floods could constitute between 0.3\% and 9.3\% of the global GDP by 2100 \cite{hinkel2014}. These effects are not linear in SLR following the same logic discussed in \cite{merel2021climate,auffhammer2018quantifying,burke2015global} for temperature changes. 

For the purpose of accurately predicting future SLR effects and formulating appropriate policies to tackle these climate change outcomes, it is essential to expand our understanding of past SLR effects \cite{tol1,novavckova2018effects}. Studies estimating the future impact of climate change in general, and SLR in particular, often rely exclusively on simulation models (e.g., \cite{RN10,RN9,RN8,RN4,RN7}) where model validation and parameter estimation are still rare \cite{hinkel2014}. As a result, empirical evidence on the SLR impact on economic growth is still missing. One exception is the study by \cite{novavckova2018effects} that uses data over a period of 30 years for the US but finds no significant effects of SLR on GDP. The authors argue that the short length of the time-series used did not suffice to identify the effects of SLR, given the slow pace and limited variability of SLR. However, one may also argue that historical SLR has had no economic impact, given the expected annual damage of coastal flood in Europe presently of €1.25 billion \cite{melet2021european} or 0.0082\% of the total EU GDP. 

Despite this minimal impact, investments in protection against SLR have already begun in Europe \cite{mcevoy2021european}, in anticipation of future damages. As slow SLR at low levels would not damage the capital stock of the economy or depreciate it faster, any economic effect of SLR at this point would be a result of investments in protection. Assuming that Europe is relatively protected against low levels of SLR \cite{mcevoy2021european} but with substantial regional variation, we argue that economies have had to reallocate resources towards less productive uses. Dikes are an expensive way to build a road or provide grazing for sheep, if this capital has no protective uses. This reallocation crowds funds away from other socially and economically useful investments, and so has a net negative impact on economic output \cite{Fankhauser2005}. However, this perceived loss does not consider the future potential damages that will occur in the absence of adaptation measures. Comparing the reallocation effect with the direct damages without adaptation measures should guide future adaptation planning. 

Following the findings of \cite{novavckova2018effects} and assuming that the protection investments in the USA are lower than the European ones \cite{magnan2023status}, we test if the absence of a SLR effect on the US GDP is due to the low amount of adaptation investments. Greater SLR would have an effect on capital, but has yet to be observed in Europe or the US. We would expect that, with greater SLR, the least adapted economies would face the highest losses but through a different mechanism, as destruction or depreciation of capital. As such, the true cost of capitalised adverse impacts of climate change can be differentiated depending on the propensity towards protection and the sea level. In a policy making context, it is important to distinguish between the cost of direct climate impacts, such as those resulting from SLR, and the costs associated with mitigation and adaptation strategies. In this context, the two are not substitutes for each other but rather complementary elements of a total assessment. The economic analysis of SLR impacts must therefore consider not only the direct costs of climate-induced changes, but also the costs and benefits of protective measures. 

In this analysis, we extend the models of \cite{dell2009temperature} and \cite{merel2021climate} on how economies react to climatic changes, such as temperature variations, to include the impact of SLR. In this context, SLR can be viewed as a defensive investment stimulus that influences GDP. Following \cite{merel2021climate}, in the short-term, GDP is affected by both the observed sea level and the deviation from the average sea level in each region. The latter term represents the immediate response to the economic shock triggered by a sea level different from what the economies are used to. Over the long-term, as regions adapt, this deviation diminishes, and GDP is primarily influenced by the actual sea level. This adaptation reflects an optimal alignment between economic actions and sea level conditions, allowing us to analyse the economic effects of SLR in both short- and long-term contexts and to assess how economies adapt to changes in sea levels. However, SLR adaptation is implicit in our analysis, as it relies on the comparison between short- and long-term outcome responses \cite{burke2015global,merel2021climate}.

To address this gap, we use a novel dataset covering a century of economic and SLR data, comparing the effects across various coastal regions in Europe. We employ data on regional SLR, specifically the \emph{relative} sea level rise (RSLR) annual dataset from the Permanent Service for Mean Sea Level (PSMSL) \cite{PSMSL_2023} as described by \cite{holgate2013new}, and data on economic growth from version 6 of the Ros\'es-Wolf database on regional GDP \cite{roses2021regional} in Europe over the past 120 years. Using these data, we identify how economic growth has been affected by past changes in sea level, such as how low SLR affects GDP through changes in the capital accumulation and productivity processes (see \textit{Methods} for identification strategy). This allows us to go beyond the simple short-term capital losses due to SLR and instead elicit the long-term effect of SLR on economic growth rates.  

Our estimates of the effects of SLR on economic growth using historical observations for the EU\&UK NUTS2 regions between 1900 and 2020, show that after 1980 SLR has already affected European regional economies. For 1 additional metre of sea level, we estimate the short-term effect on GDP to be -13.8\% and the long-term effects to be -9.6\%. We attribute this difference to adaptation effects as defined by \cite{merel2021climate} and \cite{dell2009temperature}. In the short-term, protection investments divert capital from more productive economic activities, but in the medium term (10-year period) the total effect of protection investments averages out to almost half (-7.2\%) through redistribution of resources in a dynamic setting (see \cite{dell2008climate} and Eq. \ref{reg_dynamic}). Then, extrapolating the empirical estimation beyond the highest sea level included in our dataset, we predict the regional  GDP losses of European NUTS2 regions for different SSP and RCP scenarios till 2100. Our empirically-grounded predictions indicate that GDP losses in 2100 could range between -4.6\% and -14.1\%, across the coastal EU\&UK regions in the extreme SLR scenario (SSP5-RCP8.5 High Ice) assuming BAU adaptation levels. 

\section*{Results}
\subsection*{Historical SLR and adaptation effects on GDP}

\begin{figure}[ht!]
        \centering
        \includegraphics[width=0.9\textwidth]{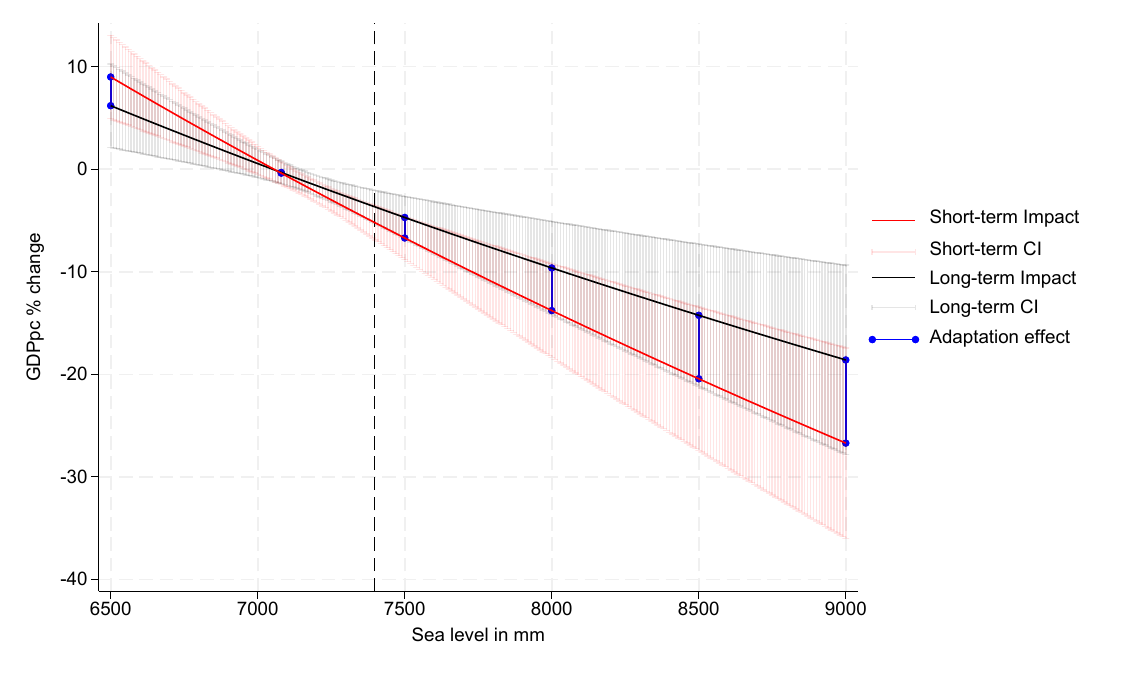}
        \caption{Short- and long-term economic impacts of sea level changes on GDPpc growth based on the adaptation model Eq. \ref{reg_base}. The blue, capped symbols with label "Adaptation effect" represent the range between the short- and long-term GDP responses to changes in sea levels. The light red range with label "Short-term CI" indicates the confidence interval for the immediate impact of sea level changes on GDP, while the grey range, labelled as "Long-term CI," displays the confidence interval for the long-term impact. The dashed line at 397mm is the maximum SLR observation in the dataset ($RSLR_{i,2020}-RSLR_{i,1900}$). Beyond this point we see a trend extrapolation of the effects.}
    \label{fig:main2}
\end{figure}

We examine the economic impacts of SLR using our estimation results depicted in Figure \ref{fig:main2} and Table \ref{tab1}, at different levels: an increase in sea level of 50 cm (7500 mm), 1 metre (8000 mm), and 2 metres (9000 mm) compared to 7000 mm, which is the Revised Local Reference (RLR) datum at each station, and is arbitrarily chosen by PSMSL to avoid negative numbers in the resulting RLR monthly and annual mean values \cite{PSMSL_2023,psmsl_2017}.

After controlling for the dynamics in GDP, and time, regional, and country effects, Table \ref{tab1} depicts the point estimates at different sea levels. The analysis separates the immediate and dynamic (lagged) effects of SLR on GDP based on \cite{dell2009temperature} and the adaptation estimation based on \cite{merel2021climate}. As expected at 7000 mm (or conventionally no SLR or no protection stimulus), there is no significant effect on GDP per capita (GDPpc). However, as the sea level increases beyond 7156 mm we have distinct negative effects of SLR ranging from -7\% to -27\% in the short term, and from -5\% and -18.6\% in the long term. The difference between the short- and long-term effects in the adaptation estimation (i.e., the adaptation effect) is indicated in Figure \ref{fig:main2}. 

At first sight, the size of the effects in Table \ref{tab1} are large; in the long term, half a metre of SLR would reduce GDP by 4.7\%. However, assuming that it would take 80 years for the sea to rise by that much, this amounts to a reduction in the growth rate of 0.06\% per year. Over the observed period of 120 years, the actualised impact of SLR on the annual growth rate ranges from -0.02\% (reduction) to 0.04\% (increase) per year, indicating that at low levels of SLR there are still positive effects on GDP growth for some regions.

\begin{table}
\centering
\caption{Point estimates of GDP per capita changes at different sea levels. For a hypothetical region, the table indicates the change in GDPpc for different sea levels, using the Dynamic estimation (Eq. \ref{reg_dynamic}) and the Adaptation estimation (main model; Eq. \ref{reg_base}).}
\begin{tabular}{lcc|cc}
\hline
\multicolumn{1}{l}{\textbf{Sea level}} & \multicolumn{1}{c}{\textbf{Immediate}} & \multicolumn{1}{c}{\textbf{Lagged}} & \multicolumn{1}{c}{\textbf{Short-term}} & \multicolumn{1}{c}{\textbf{Long-term}}  \\ 
\multicolumn{1}{l}{\textbf{in mm}} & \multicolumn{1}{c}{\textbf{effect}} & \multicolumn{1}{c}{\textbf{effect}} & \multicolumn{1}{c}{\textbf{effect}} & \multicolumn{1}{c}{\textbf{effect}}   \\ \hline
& \multicolumn{2}{c|}{Dynamic estimate} & \multicolumn{2}{c}{Adaptation estimate}   \\  \hline
6500 & 9.6\% & 6.4\% & 9.0\% & 6.2\%   \\
7000 & 0\% & 0\% & 0\% & 0\%   \\
7500 & -8.8\% & -4.3\% & -6.7\% & -4.7\%   \\
8000 & -15.7\% & -7.2\% & -13.8\% & -9.6\%  \\
8500 & -22.9\% & -10.2\% & -20.4\% & -14.2\%    \\
9000 & -29.6\% & -13.1\% & -26.7\% & -18.6\%   \\ \hline
\end{tabular}
\label{tab1}
\end{table}

In examining the SLR effects on GDP growth, we observe that current adaptation measures are adequate up to an actual sea level of 7156mm, where negative effects become statistically significant. Historical data in 2020 shows a rise of 397 mm from 1900 for some regions, yet this increase has not necessitated additional adaptations beyond what was already in place, underscoring a period where the incremental costs of further measures do not rise significantly. This can indicate a period where SLR protection investments have not reached the critical mass to affect economic output.

However, this stability is disrupted at higher levels of SLR. Our threshold analysis indicates a tipping point at the aforementioned 7156 mm, below the peak level of 7200 mm recorded in 2021. This suggests that we are currently crossing into a zone where marginal costs for SLR adaptations\textemdash through capital reallocation or loss\textemdash are on the rise. This marks a critical point where existing measures are becoming inadequate, or the economic ramifications are set to escalate disproportionately to what has been previously observed.

\begin{figure}[htp]
    \centering
    \includegraphics[width=1\linewidth]{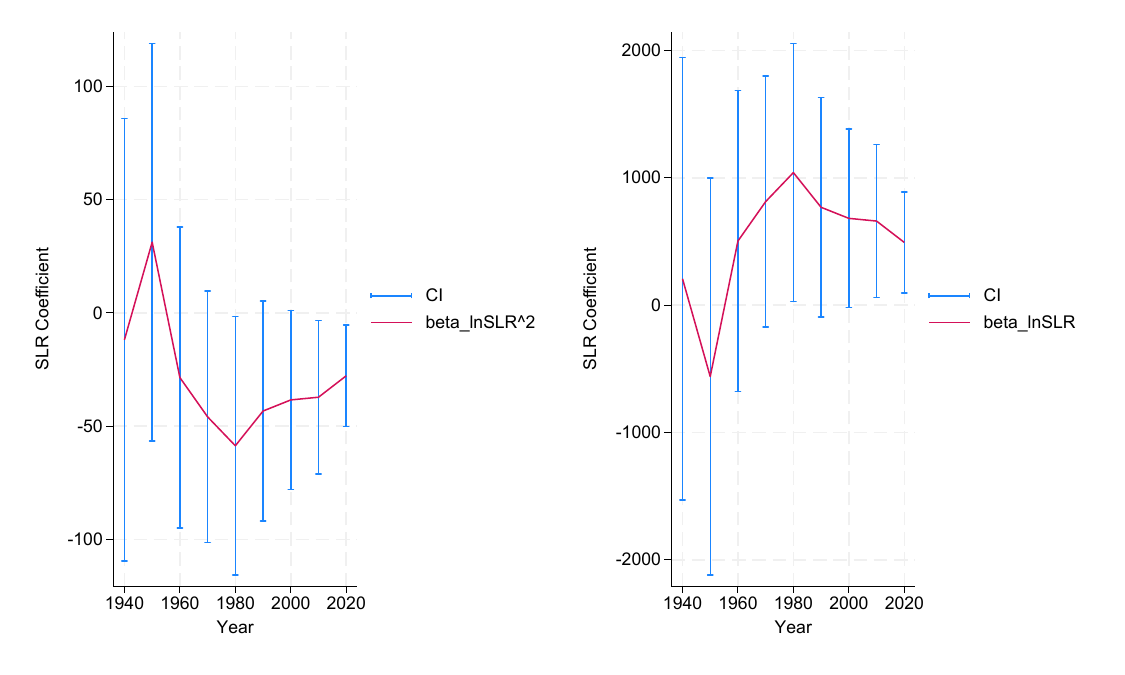}
    \caption{Rolling regression coefficients and CI at 95\% by year.}
    \label{Roll}
\end{figure}

Similarly, a rolling regression approach shows that the impact of SLR on GDP becomes statistically significant around the year 1980 (see section \textit{Extended Data} Figure \ref{Roll}). This technique, which fits a regression model to a moving window of observations, allows us to track the evolving influence of SLR on economic output over time. The results suggest that the observable economic effects of SLR post-1980 are not primarily due to capital destruction from extreme events\textemdash given the low levels of SLR we have experienced\textemdash but rather from the gradual, yet increasing, investments in adaptation. These investments towards infrastructural modifications to counteract SLR can lead to a reallocation of resources away from immediately productive uses toward long-term protective measures. While such adaptation is essential for resilience, it may not yield immediate economic gains until investments have materialised to defensive capital. This indicates that there is a cost of climate change induced by the necessary reallocation of capital and investments, even if capital is not destroyed and production is not interrupted. 

At sea levels of 7250 mm, 8000 mm, and 9000 mm, the statistically significant estimates and the difference between the short- and long-term effects imply that the economic impact of SLR on GDP per capita changes over time. This difference is the adaptation effect as described by \cite{merel2021climate}. The immediate adaptive investments or damages caused by such levels of SLR could be different from the more permanent structural changes that economies make in the long term. In this context, short-term costs could be adaptation investments, immediate disaster relief, and infrastructure repairs, while long-term costs might encompass capital changes and changes in land use. The initial loss of productive resources is depicted in the higher immediate effect, while the  economic adjustments that follow, including the materialisation of protection investments, cause the long-term effect to be smaller. 

Investments in protection are not neutral in productivity based on our estimates depicted in Figure \ref{fig:main2}. Preventive adaptation measures can initially affect GDP negatively, since productive capital is reallocated to less productive uses. We can distinguish this effect from our data since SLR has not caused any significant capital damages at the very low sea levels we have seen in Europe. However, after the investments have been transformed to protective capital and economies have reallocated their resources, the final effect of SLR becomes smaller in magnitude and can be also positive. The characteristics of this effect depend on the regional economic structure and SLR faced by each region.

\begin{figure}[ht!]
    \centering
    \includegraphics[width=\linewidth]{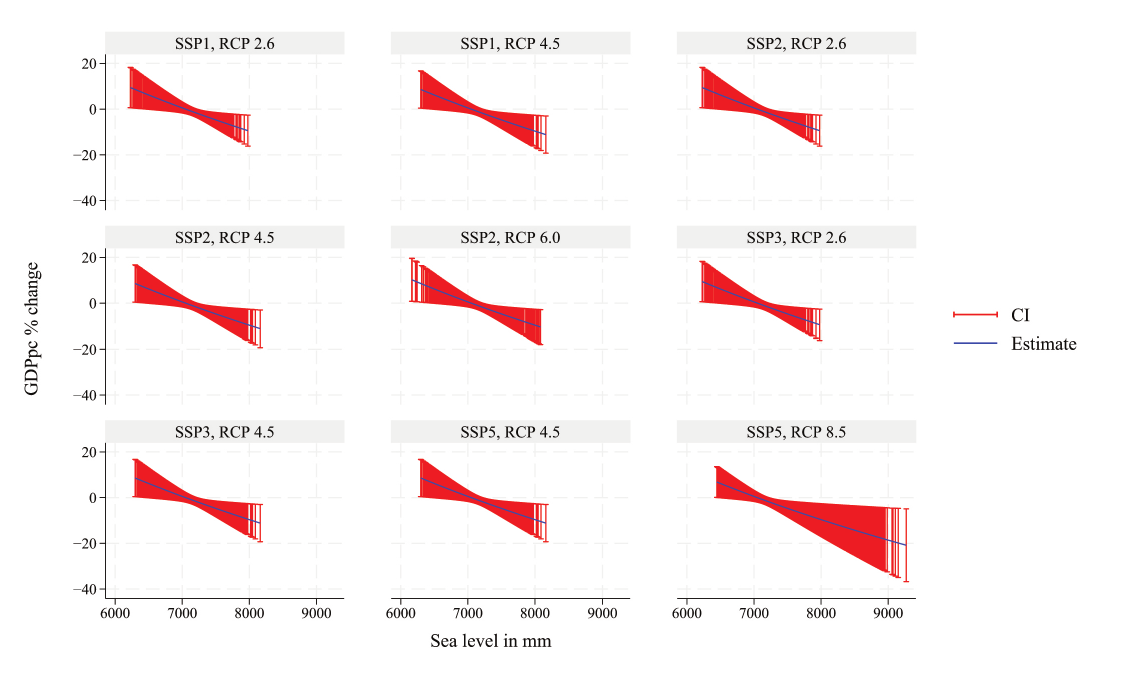}
    \caption{Projected economic impacts of SLR on GDP per capita growth, based on SSP and RCP projection scenarios from Eq. \ref{reg_base1} for all regions.}
    \label{fig:SSPRCP}
\end{figure}

\subsection*{Regional economic projections on higher SLR levels}

\begin{figure}[ht!]
    \centering
    \begin{subfigure}{.49\textwidth}
        \centering
        \includegraphics[width=\textwidth]{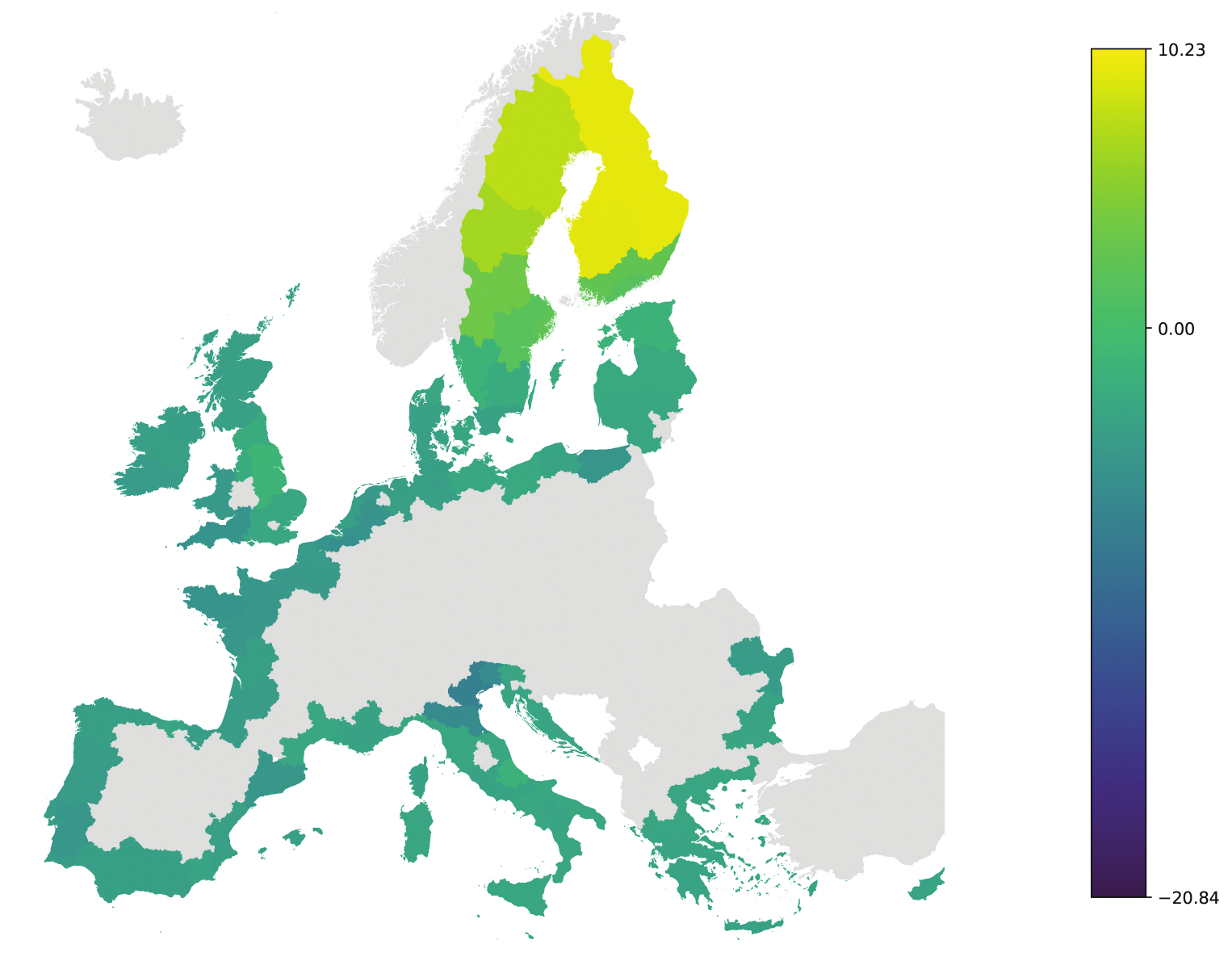} 
        \caption{SSP1-RCP2.6-Low Ice}
    \end{subfigure}\hfill
    \begin{subfigure}{.49\textwidth}
        \centering
        \includegraphics[width=\textwidth]{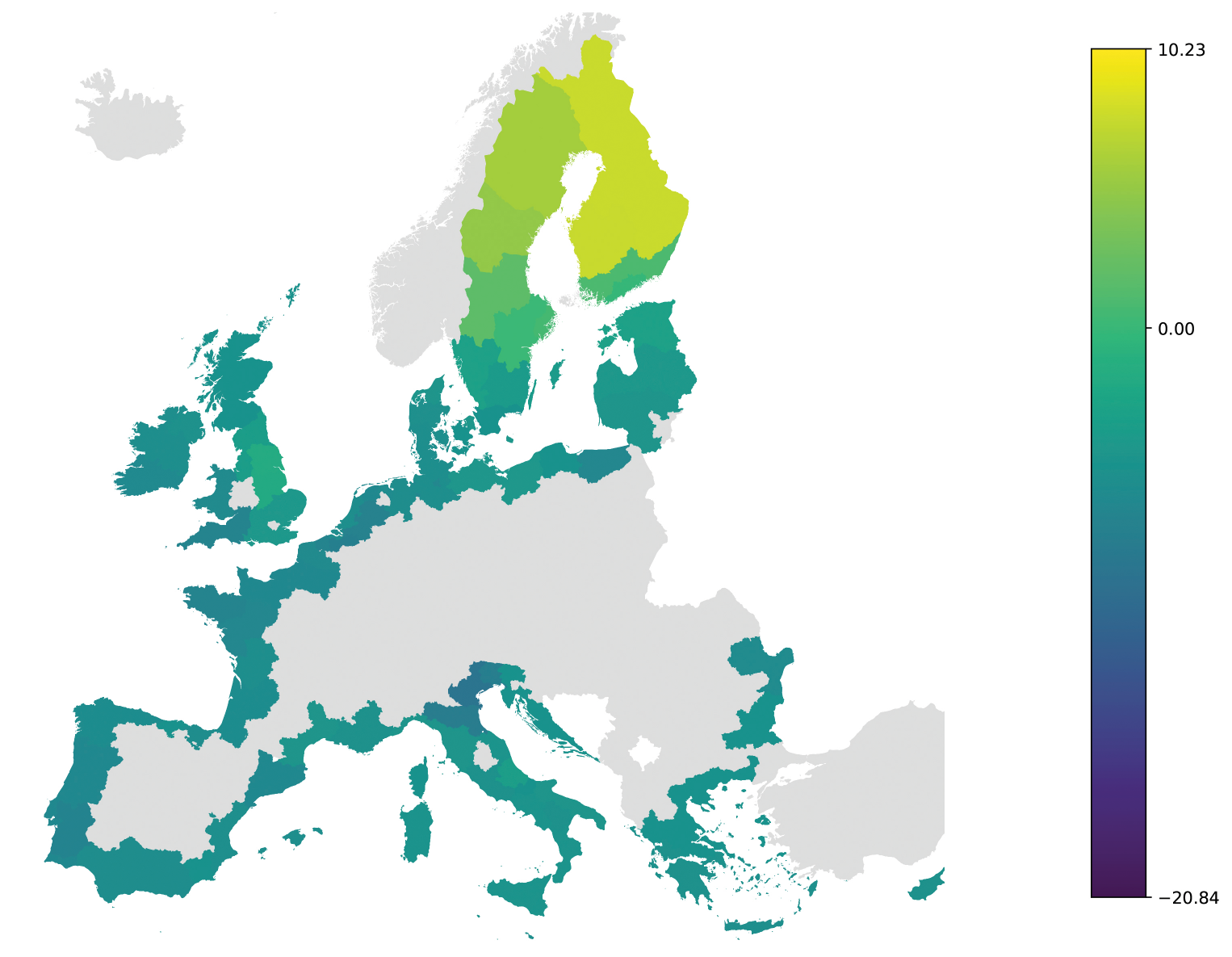} 
        \caption{SSP2-RCP4.5-Medium Ice}
    \end{subfigure}

    \vspace{1cm}

    \begin{subfigure}{.50\textwidth}
        \centering
        \includegraphics[width=\textwidth]{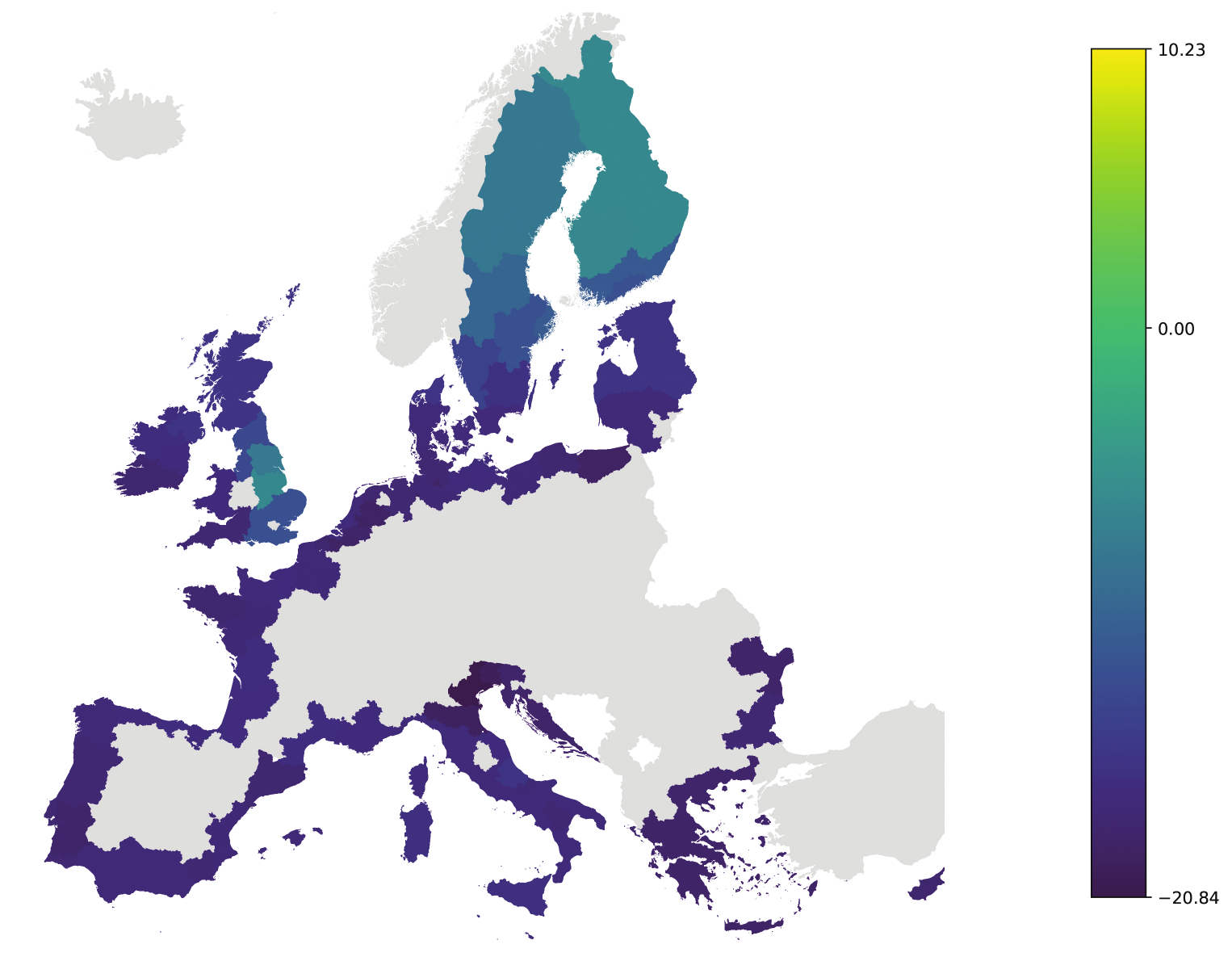} 
        \caption{SSP5-RCP8.5-High end Ice}
    \end{subfigure}

    \caption{Regional projections of GDP losses in Europe under different combinations of SSP, RCP and Ice scenarios. From yellow (higher) till green (lower), represent the positive GDP per capita change, and moving from green to blue indicates higher GDP per capita losses, based on the SLR and population projected for each region in every scenario for 2100 based on Eq. \ref{reg_base1}.}
    \label{fig:GDP}
\end{figure}

To complement the aggregated assessments of the relationship between GDP and SLR for all European regions, we visualise the regional pattern of the projected impacts of different SLR scenarios on the average GDP growth rate, focusing on the coastal regions (Figures \ref{fig:SSPRCP} \& \ref{fig:GDP}). Notably, using the relationships revealed from historic trends, we explore how GDP per capita of coastal regions may change under different combinations of SSP and RCP scenarios. 

Using the COACCH project's \cite{lincked2} SLR results from the DIVA modelling framework \cite{hinkel2014coastal,lincke2021coastal} at the NUTS2 level, we estimate the GDP changes for each NUTS2 region based on the region-specific population and SLR projections under different scenarios. As expected, the adverse consequences of SLR on GDP increase as one goes from the low- to medium- and high-impact scenarios, corresponding to three combinations of socioeconomic (SSP), climate (RCP) and ice sheet decline (Ice) scenarios: SSP1-RCP2.6-Low Ice, SSP2-RCP4.5-Medium Ice, and SSP5-RCP8.5-High-end Ice in Figure \ref{fig:GDP}. Additionally, we perform an extensive sensitivity analysis across all other combinations (see \textit{Extended Data} section for details).

\begin{table}[!ht]
    \centering
    \caption{Highest and lowest projected cumulative GDP per capita growth changes in 2100 by NUTS2 region for SSP5-RCP8.5-High-end Ice, SSP2-RCP 4.5-Medium Ice, SSP1-RCP 2.6-Low Ice (based on estimating Eq. \ref{reg_base} using the long-term effects).}
    \begin{tabular}{llcllc}
    \hline
        \textbf{Nuts2} & \textbf{Name} & \textbf{Change} & \textbf{Nuts2} & \textbf{Name} & \textbf{Change} \\ \hline
        \multicolumn{3}{c}{\textbf{SSP5-RCP8.5}} & \multicolumn{3}{c}{\textbf{SSP2-RCP4.5}} \\
        ITH3 & Veneto & -20.8 & ITH3 & Veneto & -7.3 \\ 
        ITH4 & Friuli Venezia Giulia & -19.9 & ITH5 &  Emilia-Romagna & -6.1 \\ 
        PT30 & R. A. da Madeira & -19.7 & ITH4 & Friuli Venezia Giulia & -6.0 \\ 
        ITH5 &  Emilia-Romagna & -19.5 & BE23 & West-Vlaanderen & -5.6 \\ 
        PT20 & R. A. dos Açores & -19.5 & BE21 & Antwerpen & -5.6 \\ 
        SE32 & Mellersta Norrland & -8.2 & SE31 & Norra Mellansverige & 3.6 \\ 
        SE33 & Övre Norrland & -8.1 & SE32 & Mellersta Norrland & 6.5 \\ 
        FI19 & Länsi-Suomi & -6.5 & SE33 & Övre Norrland & 7.6 \\ 
        UKF & East Midlands  & -6.4 & FI19 & Länsi-Suomi & 9.3 \\ 
        FI1D & Pohjois-ja Itä-Suomi & -6.3 & FI1D & Pohjois-ja Itä-Suomi & 9.5 \\ \hline
        \textbf{Nuts2} & \textbf{Name} & \textbf{Change} & ~ & ~ & ~ \\ \hline
        \multicolumn{3}{c}{\textbf{SSP1-RCP2.6}} & \multicolumn{3}{c}{\textbf{}} \\
        ITH3 & Veneto & -8.9 & ~ & ~ & ~ \\ 
        ITH5 &  Emilia-Romagna & -7.8 & ~ & ~ & ~ \\ 
        ITH4 & Friuli Venezia Giulia & -7.7 & ~ & ~ & ~ \\ 
        BE23 & West-Vlaanderen & -7.5 & ~ & ~ & ~ \\ 
        BE21 & Antwerpen & -7.5 & ~ & ~ & ~ \\ 
        SE31 & Norra Mellansverige & 1.9 & ~ & ~ & ~ \\ 
        SE32 & Mellersta Norrland & 4.9 & ~ & ~ & ~ \\ 
        SE33 & Östra Sverige & 5.9 & ~ & ~ & ~ \\ 
        FI19 & Länsi-Suomi & 7.6 & ~ & ~ & ~ \\ 
        FI1D & Åland & 7.7 \\ \hline
    \end{tabular}
    \label{max}
\end{table}

The SSP1-RCP2.6-Low Ice pathway (Figure \ref{fig:GDP}.a and Table \ref{max} top-left) envisions a future with low population growth and rapid shifts towards a more sustainable mode of development. Under this scenario, there is likely to be less strain on resources due to a smaller global population. In combination with RCP2.6, this scenario anticipates a significant reduction in greenhouse gas emissions and thus a minimal sea level rise. The implications for GDP are likely to be less severe because less land would be at risk of inundation. Still, even in this low impact scenario, our results indicate GDP changes of approximately -4.2\% on average by 2100. Given our identification strategy, this effect can be attributed to the investment reallocation resulting from the SLR stimulus. In this scenario, there are also positive estimates in the long-term, especially in Scandinavian regions where we see a land gain. 

In the SSP2-RCP4.5-Medium Ice pathway (Figure \ref{fig:GDP}.b and Table \ref{max} top-right), current socioeconomic trends\textemdash including moderate population growth\textemdash continue (SSP2), coupled with a moderate increase in greenhouse gas emissions that stabilise before 2100. These climatic conditions drive a moderate level of SLR. In this medium-impact scenario combination, our results reveal GDP changes of approximately -5\% on average across European regions by 2100 These potential impacts on GDP are 20\% higher on average than under the low impact scenario (comparing Figures \ref{fig:GDP}.a and \ref{fig:GDP}.b). Similarly to the low -impact pathway, ITH3-Veneto, ITH5-Emilia-Romagna, ITH4-Friuli-Venezia Giulia, and the Belgian coast face the highest losses (approximately -6\% by 2100). However, positive effects still exist in this scenario caused by the land gain in the northern regions of Scandinavia.

Finally, the SSP5-RCP8.5-High-end Ice pathway (Figure \ref{fig:GDP}.c and Table \ref{max} bottom) presents a world with significant population growth and high fossil-fuel-driven economic growth, which would likely lead to high demand for resources and substantial strain on the environment. The high-end climate RCP8.5 scenario assumes a continued rise in greenhouse gas emissions throughout the 21st century, which leads to major SLR. This combination has a profound impact on GDP due to the large population increase and significant economic infrastructure at risk from rising sea levels. Specifically, this high-end pathway leads to more than 9\% reduction in GDP on average for coastal regions in Europe by 2100; these GDP losses double compared to the low-impact pathway (compare Figures \ref{fig:GDP}.a and \ref{fig:GDP}.c). ITH3-Veneto experiences the highest losses at -21\% of regional GDP, followed by ITH4-Friuli-Venezia Giulia (-20\%). In this scenario, all regions have GDP per capita losses by 2100.  

\section*{Discussion} 
Climate-induced SLR is already damaging coastal economies, and its adverse effects will only increase. Yet, eliciting empirical evidence on the effect of such a slow-mowing variable as SLR on economic growth has not been straightforward. Hence, most estimates of the costs of SLR are based on model assessments, and often include only the short-term direct damages, omitting the long-term impact of SLR on GDP growth. Using the novel combination of data on both GDP and SLR at the regional scale in Europe over 120 years, we fill this gap, offering unique empirical evidence on how SLR affects economic growth through the reallocation of investments and capital.

Previous literature reported more modest effects of SLR, where for example coastal flood damage under higher-end SLR (1.3 m) would be approximately 4\% of world GDP annually\cite{oecd2019responding} (USD 50 trillion annually, without adaptation), and GDP losses from SLR in Northern Europe were estimated between 4.4\%-5.3\% (SSP2 vs. SSP5) \cite{parrado2020fiscal}. In contrast, our analysis estimates a 9.6\% reduction in GDP at 1.3 metres of SLR, double the OECD estimate. However, the mechanism through which SLR affects GDP is different: earlier estimates are comparative static, whereas we consider the impact on cumulative growth. Given our estimating strategy, we predict a 2.9\% long-term reduction in GDP per capita on average by 2100 for SSP1-RCP2.6-Low Ice, 4.7\% on average for SSP2-RCP4.5-Medium Ice and a 10.4\% in 2100 for SSP5-RCP8.5-High Ice pathway.

Compared to the only empirical analysis linking SLR to economic growth \cite{novavckova2018effects}, our study covers a period that is four times as long, 120 years of observations. This turns out to be essential to capture a statistically significant effect of such a slow-moving variable as SLR on GDP, which is, in turn, subject to the influence of many drivers besides climate change. Our additional assumption that the insignificant effects in \cite{novavckova2018effects} are due to the historical level of investments in adaptation, is rejected. We find no significant effect when constraining our analysis to the period 1980-2020, so we conclude that indeed the series length is the most important factor in revealing the effects of SLR on GDP.

While the literature has anticipated non-marginal effects of climate adversities, including SLR, on GDP \cite{berrang2021systematic}, to our knowledge, this study is the first to empirically estimate such a relationship for SLR and GDP while differentiating between short- and long-term effects. We show that the initial reduction in GDP persists in the future but is significantly diminished by adaptation. We observe the negative effect of climate change in the economy caused by the reallocation of investment to less productive capital accumulation. However, these 'sub-optimal' capital investments are of course necessary to protect against the massive damages SLR could bring in the future.

However, capital reallocation is not the only path through which SLR can affect GDP within on our identification strategy. The uncertainty associated with SLR could lead to an increased risk aversion among investors, businesses and the public sector. This could result in reduced investment, lower risk-taking, and a general slowdown in economic activities in regions for which SLR has been already observed or where traditional adaptation via seawalls, dikes and beach nourishment is not attainable. Furthermore, SLR can alter beaches, increase flooding, and affect the overall attractiveness of coastal regions as tourist destinations \cite{Hamilton2007}. In turn, this can have a negative effect on GDP. Another possible path that can cause this effect is infrastructure degradation \cite{abdelhafez2022hidden}. The gradual effect of SLR on existing infrastructure (like erosion, increased maintenance, etc.) can lead to higher public and private expenditure on infrastructure upkeep, further reducing the resources available for other economic activities.

This work is not without limitations and can be expanded in several directions. Adaptation will be crucial to curb the impacts of SLR, and the ability to take this action highly depends on the economic capacity of a region\cite{Yohe2002}, and hence on GDP growth. Given this feedback between economic development and adaptive capacity of regional economies, the current ex-ante analysis assumes adaptation to be constant at 2020 levels. Future work could focus on various regional-level scenarios of adaptation that are contingent on the longitudinal projections of GDP.

In addition, GDP as a measure of economic development suffers from multiple flaws, already extensively discussed in the literature \cite{stiglitz2009report,chancel2014beyond, EuropeanCommission2021}. Specifically to disasters, the standard method to estimate GDP assumes that investments in recovery and protection add value to the economy, and thus are positive contributors to GDP. Our results based on historical data show the opposite: a statistically significant reduction of GDP growth rate at high levels of SLR. For low levels of SLR, both the growth and level effect of SLR on GDP can either be positive or negative, indicating the complexity through which SLR affects the economy in the short- and long-term.

Moreover, our analysis focuses on European regions only, where we were able to find data over such long timescales at a fine regional resolution. It is important to extend the analysis to other regions in future work, especially as SLR effects are so region-specific. For example, \cite{roses2018economic} reports USA state-level data on GDP from 1880 to 2010. However, the USA data are measured differently to the European ones, creating inconsistencies \cite{roses2018economic}, and hence could not be included in the current analysis. However, a direct comparison with the results of \cite{novavckova2018effects} can be made since the statistical estimation models are identical. 


Lastly, our findings offer a fresh perspective on coastal retreat as a strategic response to SLR. In certain cases, strategically relocating assets and capital from climate-sensitive areas might be more efficient than protecting them, as accelerating costs of protection diverge resources away from 'productive' investments, as our estimated negative effects reveal. Since inaction towards SLR will be more costly \cite{vousdoukas2018climatic} by 2100 through the depreciation of capital, the discourse should move away from protection vs. no-protection, to protection vs. \emph{retreat} by identifying the cases where the retreat is preferable and in what form \cite{lincke2021coastal}. Coastal setback zones - areas along coastlines where further development is restricted or prohibited - could also be an alternative. These areas can reduce the impact of coastal hazards by 50\% in the majority of EU countries by 2100 \cite{wolff2023setback}. However, the total economic effects of coastal setback zones have not yet been assessed.

\noindent

\section*{Methods}\label{sec1}

\subsection*{Identification strategy and estimation model}

Our estimation strategy follows \cite{dell2009temperature}, \cite{mohaddes2023climate} and particularly \cite{merel2021climate}. 
We assume that GDPpc is a maximised outcome that depends on a climatic effect $x_{i,t} \in X_{i}$ $\subseteq$ $\Re$, a set of actions $\xi_{i} \in \Re^{L}$ taken by economic agents, and an error term $\epsilon_{i,t}$:

\begin{align} 
 lnGDPpc_{it} = \phi_{i}(x_{i,t}, \xi_{i}, \epsilon_{i,t})
\end{align}

The climatic effect in \cite{dell2009temperature, mohaddes2023climate, merel2021climate} is temperature. We adapt this to SLR assuming a quadratic functional form that captures non-monotonicities and non-linearities in the SLR-outcome relationship.

Beyond that, following \cite{merel2021climate} introduce a SLR penalty $(x_{i,t} - \omega_{i}(\xi_{i}))^{2}$, indicating that conditional on SLR $x_{i,t}$, the outcome is maximised if and only if $x_{i,t} = \omega_{i}(\xi_{i})$ where $\xi_{i}$ is a set of actions, leading us to:

\begin{align} \label{reg_baseX}
    lnGDPpc_{it} = & \alpha_{i,t} + \beta_{1}lnSLR_{i,t}+ \beta_{2}lnSLR_{i,t}^{2}+ \beta_{3}((x_{i,t} - \omega_{i}(\xi_{i}))^{2}  + \epsilon_{it}
\end{align}

Or in growth form instead of level that now includes the convergence tern $lnGDPpc_{i,t-10}$ and speed of convergence $\theta$:

\begin{align} \label{reg_base}
    \Delta lnGDPpc_{i,t} = & \alpha_{i,t} + \beta_{1}lnSLR_{i,t}+ \beta_{2}lnSLR_{i,t}^{2}  + \beta_{3}((x_{i,t} - \omega_{i}(\xi_{i}))^{2}  + \theta lnGDPpc_{i,t-10} + \epsilon_{it}
\end{align}

Based on Equation \ref{reg_base}, we get the expected values of outcome in the long term where $lnSLR_{i,t} = \Bar{lnSLR_{i}}$ as:

\begin{align} \label{reg_base1}
    \mathbb{E} [\Delta lnGDPpc_{i}^{LT}|\alpha_{i},lnSLR_{i}]= & \alpha_{i,t} + \beta_{1}lnSLR_{i,t}+ \beta_{2}lnSLR_{i,t}^{2}
\end{align}

Here in the hypothetical long-term all actions $\xi_{i}$ can be varied in response to changes SLR, the penalty vanishes. The implied short-term response to SLR, given the penalty is:

\begin{align} \label{reg_base2}
    \mathbb{E} [\Delta lnGDPpc_{i}^{ST}|\alpha_{i},\mu_{i},lnSLR_{i}]= & \alpha_{i,t} + \beta_{1}lnSLR_{i,t}+ \beta_{2}lnSLR_{i,t}^{2}  + \beta_{3} (lnSLR_{i,t} - \Bar{lnSLR_{i}})^{2} 
\end{align}

Based on Eq. \ref{reg_base1} and Eq. \ref{reg_base2}, the adaptation effect is the difference between the short- and long-term effects.

Following \cite{dell2008climate} but on a quadratic function, we also estimate a dynamic structure separately in order to assess the lagged effect of SLR on GDP per capita, separating the growth from the level effect of SLR on GDP: 


\begin{align} \label{reg_dynamic}
   \Delta lnGDPpc_{i,t} = & \alpha_{i,t} + \theta_{1}lnGDPpc_{i,t-10} + (\beta_{1} +\gamma_{1})lnSLR_{i,t} + (\beta_{2} +\gamma_{2})lnSLR^{2}_{i,t}  \nonumber \\
    &  - \beta_{1} lnSLR_{i,t-10} - \beta_{2} lnSLR^{2}_{i,t-10}  + FEs + u_{i,t}
\end{align}

Where the level effect enters through $\beta$ and the growth effects through $\gamma$.

Country/time fixed effects ($FEs$) are used to control for time-varying factors that affect all NUTS2 in the same country in the same way. These factors include global economic trends, technological progress, common shocks like WWI, WWII, the creation of the EU, climate change other than sea level rise, or any other global events that might influence GDP growth across all countries in the dataset. These, on top of regional $FEs$ to control for the time invariant differences between the regions.

\subsubsection*{Data used}
\begin{figure}
    \centering
    \includegraphics[width=0.6\linewidth]{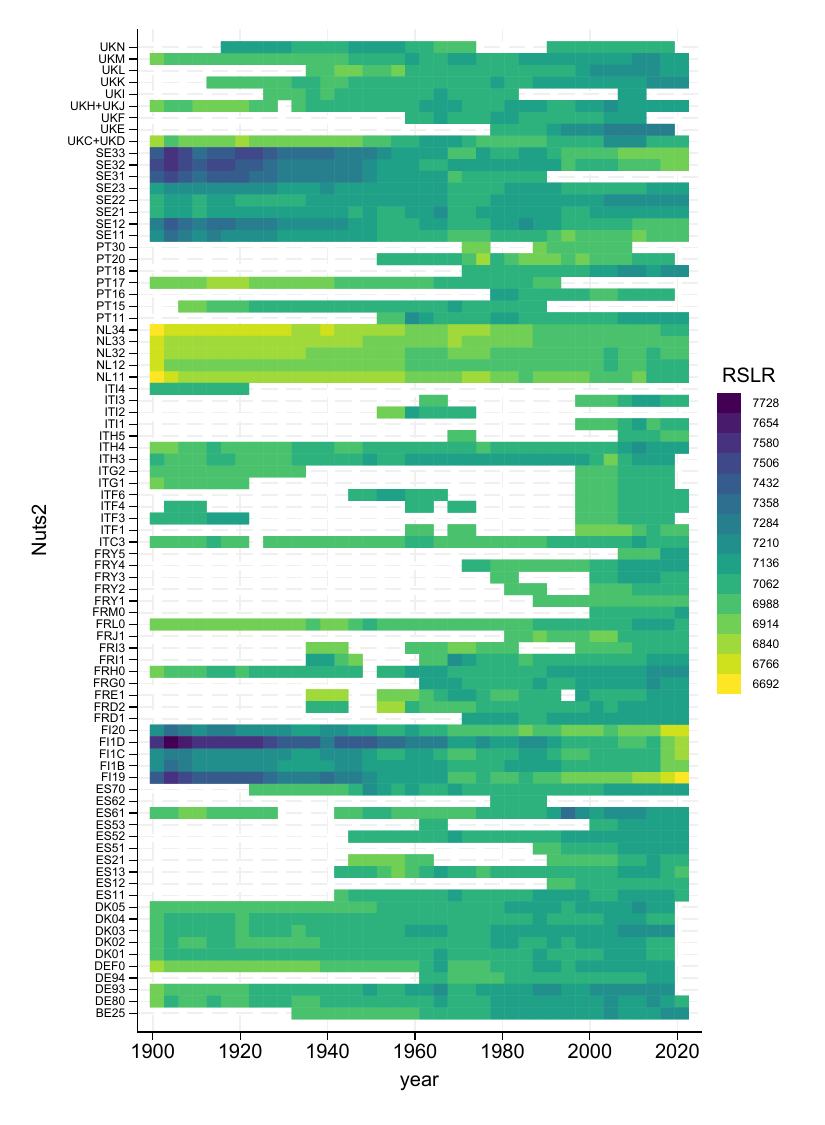}
\caption{Sea level measurement in mm derived from the Permanent Service for Mean Sea Level (PSMSL) RLR annual dataset\cite{PSMSL_2023} as described by \cite{holgate2013new} for a subset of regions}     
\label{fig:Data}
\end{figure}

The tide gauge data representing SLR are derived from the Permanent Service for Mean Sea Level (PSMSL) RLR annual dataset \cite{PSMSL_2023} as described by \cite{holgate2013new}. The SLR data are then matched to the NUTS2 regions where is station is placed. If a NUTS2 includes more than one station the values per year are averaged. If a NUTS2 regions does not include a station, SLR for that NUTS2 is indicated as missing values and it is excluded from the analysis, contrary to \cite{novavckova2018effects} that use the values from the closest station. In the context of SLR, this is problematic because it assumes relative small variation in SLR in neighbouring regions, an assumption that is particularly problematic in seismically active areas. In our estimates, we only use the SLR values we are sure that have been observed by each region. In our analysis, the RLR annual dataset is translated to metres of SLR by subtracting the values for each station by its benchmark. A detailed depiction of the SLR data between 1900 and 2020 is included in Figure \ref{fig:Data} and Table \ref{tab:Decr}. As described by the PSMSL documentation \cite{psmsl_2017}, the RLR datum at each station is defined to be approximately 7000 mm below mean sea level, with this arbitrary choice made many years ago in order to avoid negative numbers in the resulting RLR monthly and annual mean values. As such we assume a PSMSL RLR annual measurement of 8000 mm to be approximately 1 metre of SLR.

Data on historical GDP and population for 172 NUTS2 regions (including aggregations) between 1900 and 2015 for Europe is derived from the Ros\'es-Wolf database on regional GDP version 6 \cite{roses2021regional}. Their basic methodology is described in detail in \cite{roses2018economic}. This dataset is then expanded to 2020 using the growth rate of each NUTS2 region derived from EUROSTAT \cite{Eurostat_2023a,Eurostat_GDP_2023} for population and GDP respectively.

\subsubsection*{Estimation sensitivity}
The main estimation results combined with a series of different specifications are presented in Table \ref{tab:Regressions}. 

\begin{table}[ht!]
\caption{Basic descriptive statistics}
    \centering
    \label{tab:Decr}
    \begin{tabular}{lccccc}
    \hline
        \textbf{} & \textbf{N} & \textbf{mean} & \textbf{SD} & \textbf{min} & \textbf{max} \\ \hline
        \textit{GDP(2011)} & 2,230 & 41,680 & 70,045 & 205 & 858,205 \\ 
        \textit{Sea Level*} & 15,969 & 7,007 & 105 & 6,651 & 7,765 \\ 
        \textit{Years} & 120 & ~ & ~ & 1900 & 2020 \\ \hline
        \textit{No. of} & 173 & 79 Coastal & 94 Inland & ~ & ~ \\ 
        \textit{NUTS2} & ~ & -used & -not used & ~ & ~ \\ \hline

    \end{tabular} \\
    *Sea level data for all European regions, containing the excluded from the \cite{roses2021regional} dataset.
\end{table}

\begin{table}[!ht]
\caption{Regression results for all models}
\label{tab:Regressions}
    \centering
    \begin{tabular}{lcccccc}
    \hline
       $\Delta ln(GDPpc)=$ & Adaptation & Dynamic & Linear & 1980-2020 & FEs 1 & FEs 2 \\ 
        ~ & model Eq. \ref{reg_base} & model Eq. \ref{reg_dynamic} & model & model Eq. \ref{reg_base} & ~ & ~ \\ \hline
        $lnSLR$ & 675** & 1,232*** & -0.888 & -459 & 1,863*** & 192 \\
        ~ & (-279) & (-279) & (0.820) & (-625) & (-432) & (NE) \\ 
        $lnSLR_{t-10}$ & ~ & -758*** & ~ & ~ & ~ & ~ \\ 
        ~ & ~ & (-281) & ~ & ~ & ~ & ~ \\ 
        $lnSLR^{2}$ & -38** & -69*** & ~ & 26 & -105*** & -11 \\ 
        ~ & (-16) & (-16) & ~ & (-35) & (-24) & (NE) \\ 
        $lnSLR_{t-10}^{2}$ & ~ & 43*** & ~ & ~ & ~ & ~ \\ 
        ~ & ~ & (-16) & ~ & ~ & ~ & ~ \\ 
        $lnGDPpc_{t-10}$ & -0.475*** & -0.198*** & -0.469*** & -0.930*** & -0.049*** & -0.167 \\ 
        ~ & (0.062) & (0.024) & (0.060) & (0.139) & (0.009) & (NE) \\ 
        $Penalty$ & -33** & ~ & -66*** & -3.660 & -41 & -32 \\ 
        ~ & (-15) & ~ & (-18) & (-26) & (-33) & (NE) \\
        $Constant$ & -2,989** & -2,095*** & 9.240 & 2,042 & -8,261*** & -845 \\ 
        ~ & (-1,240) & (-791) & (7.281) & (2,770) & (-1,916) & (NE) \\ \hline
        \textbf{Country-Year FEs} & Yes & Yes & Yes & Yes & No & Yes \\ 
        \textbf{Region FEs} & Yes & No & Yes & Yes & Yes & No \\ 
        Observations & 629 & 629 & 629 & 307 & 629 & 629 \\ 
        R-squared & 0.743 & 0.715 & 0.741 & 0.887 & 0.142 & 0.693 \\ \hline
    \end{tabular} \\
    $Penalty=(lnSLR_{i,t}-\Bar{lnSLR_{i}})^{2}$, Robust standard errors in parentheses, clustered by NUTS2 and country/year. 
        $*** p<0.01, ** p<0.05, * p<0.1$, NE: Not estimated
\end{table}

\subsection*{Future SLR and population projections}
The projections of future coastal exposure and impacts are based on the NUTS2 data on population and SLR for different combinations of SSPs and RCPs as derived from the DIVA model \cite{hinkel2014coastal,lincke2021coastal} for the period 2025-2100. Impact and cost calculations are based on 12,148 coastal segments, as defined in the DINAS-COAST database \cite{vafeidis2008new}. These coastal segments model specific coastal areas with homogenous bio-physical and socio-economic characteristics. To determine local relative changes in sea level, DIVA supplements climate-induced SLR data with information on glacial-isostatic adjustments \cite{peltier2004global} and subsidence data for coastal segments linked to river deltas \cite{nicholls2021global}.

Following \cite{bachner2022macroeconomic}, our study also employs the RCP-SSP combination scenario framework. Representative Concentration Pathways (RCPs) are effectively the trajectories of greenhouse gas emissions over time. Four RCPs are commonly used in the modelling community, namely RCP2.6, RCP4.5, RCP6.0, and RCP8.5. As an example, RCP4.5 describes a scenario with moderate greenhouse gas emissions, leading to a radiative forcing of 4.5 $W/m^{2}$ by 2100. This scenario is likely to result in a 2.1-3.5°C rise in global mean surface temperature by 2100, compared to the 1850-1900 period. The RCP that projects the most significant radiative forcing by 2100 is RCP8.5. Under this scenario, the global mean temperature is likely to rise by 3.3-5.7°C by 2100 (compared to the 1850-1900 baseline) and the mean sea-level rise is likely to be between 0.8 and 1.2m by 2100 (relative to 1900 sea levels). 

The Shared Socioeconomic Pathways (SSPs) are scenarios about how the social, economic, and political factors that influence climate change might evolve in the future. Here is a short description for each of the five SSPs \cite{o2017roads}: SSP1: This pathway describes a world that steadily shifts towards a more sustainable path, emphasising more inclusive development that respects perceived environmental boundaries; SSP2: This pathway outlines a world where trends broadly follow their historical patterns, with uneven development and slowly decreasing fossil fuel dependency; SSP3: This pathway depicts a world of increasing nationalistic and protectionist tendencies, with an emphasis on self-sufficiency, security, and local identities leading to fragmented development; SSP4 This pathway describes a highly unequal world, where the well-off are isolated from the world's poor, who are exposed to the harmful impacts of environmental changes and have limited access to education and health care; SSP5: This pathway outlines a world with strong faith in competitive markets, innovation and participatory societies, where the push for economic and social development is coupled with exploitation of abundant fossil fuel resources, leading to high greenhouse gas emissions.

In our analysis we focus on three scenario combinations:
\begin{enumerate}
\item RCP8.5-SSP5 with a high-end ice melting assumption with 0.5 m (0.53 m) SLR by 2050 and 1.77 m (1.83 m) by 2100) (low-likelihood, high-impact\cite{IPCC_2021};
\item RCP4.5-SSP2 with medium ice melting, as a middle-of-the road scenario (but still not fully compatible with the Paris targets) with 0.17 m (0.20 m) SLR by 2050 and 0.45 m (0.50 m) SLR by 2100;
\item RCP2.6-SSP1 with low ice melting, as a optimistic scenario . It is anticipated that it will produce a mean of less than 2ºC warming by 2100 with 0.14 m (0.17 m) SLR by 2050 and 0.26 m (0.55 m) SLR by 2100.
\end{enumerate}

However, all combination projections are available in the extended data section.

\section*{Extended data}

\clearpage
\bibliography{SLR_GDP}

\begin{thebibliography}{10}
\urlstyle{rm}
\expandafter\ifx\csname url\endcsname\relax
  \def\url#1{\texttt{#1}}\fi
\expandafter\ifx\csname urlprefix\endcsname\relax\def\urlprefix{URL }\fi
\expandafter\ifx\csname doiprefix\endcsname\relax\def\doiprefix{DOI: }\fi
\providecommand{\bibinfo}[2]{#2}
\providecommand{\eprint}[2][]{\url{#2}}

\bibitem{mcgranahan2007rising}
\bibinfo{author}{McGranahan, G.}, \bibinfo{author}{Balk, D.} \& \bibinfo{author}{Anderson, B.}
\newblock \bibinfo{journal}{\bibinfo{title}{The rising tide: assessing the risks of climate change and human settlements in low elevation coastal zones}}.
\newblock {\emph{\JournalTitle{Environment and urbanization}}} \textbf{\bibinfo{volume}{19}}, \bibinfo{pages}{17--37} (\bibinfo{year}{2007}).

\bibitem{smith2011we}
\bibinfo{author}{Smith, K.}
\newblock \bibinfo{journal}{\bibinfo{title}{We are seven billion}}.
\newblock {\emph{\JournalTitle{Nature climate change}}} \textbf{\bibinfo{volume}{1}}, \bibinfo{pages}{331--335} (\bibinfo{year}{2011}).

\bibitem{copernicus2023}
\bibinfo{author}{{Copernicus Land Monitoring Service}}.
\newblock \bibinfo{title}{Coastal zones}.
\newblock \bibinfo{howpublished}{\url{https://land.copernicus.eu/local/coastal-zones}} (\bibinfo{year}{2023}).
\newblock \bibinfo{note}{Accessed: 2023-06-17}.

\bibitem{bosello2012}
\bibinfo{author}{Bosello, F.}, \bibinfo{author}{Nicholls, R.~J.}, \bibinfo{author}{Richards, J.}, \bibinfo{author}{Roson, R.} \& \bibinfo{author}{Tol, R.~S.}
\newblock \bibinfo{journal}{\bibinfo{title}{Economic impacts of climate change in europe: sea-level rise}}.
\newblock {\emph{\JournalTitle{Climatic change}}} \textbf{\bibinfo{volume}{112}}, \bibinfo{pages}{63--81} (\bibinfo{year}{2012}).

\bibitem{parrado2020fiscal}
\bibinfo{author}{Parrado, R.} \emph{et~al.}
\newblock \bibinfo{journal}{\bibinfo{title}{Fiscal effects and the potential implications on economic growth of sea-level rise impacts and coastal zone protection}}.
\newblock {\emph{\JournalTitle{Climatic Change}}} \textbf{\bibinfo{volume}{160}}, \bibinfo{pages}{283--302} (\bibinfo{year}{2020}).

\bibitem{tol1}
\bibinfo{author}{Tol, R. S.~J.}
\newblock \bibinfo{journal}{\bibinfo{title}{The economic effects of climate change}}.
\newblock {\emph{\JournalTitle{Journal of economic perspectives}}} \textbf{\bibinfo{volume}{23}}, \bibinfo{pages}{29--51} (\bibinfo{year}{2009}).

\bibitem{vousdoukas2018climatic}
\bibinfo{author}{Vousdoukas, M.~I.} \emph{et~al.}
\newblock \bibinfo{journal}{\bibinfo{title}{Climatic and socioeconomic controls of future coastal flood risk in europe}}.
\newblock {\emph{\JournalTitle{Nature Climate Change}}} \textbf{\bibinfo{volume}{8}}, \bibinfo{pages}{776--780} (\bibinfo{year}{2018}).

\bibitem{ciscar2014climate}
\bibinfo{author}{Ciscar, J.-C.} \emph{et~al.}
\newblock \bibinfo{title}{Climate impacts in europe-the jrc peseta ii project} (\bibinfo{year}{2014}).

\bibitem{schinko2020economy}
\bibinfo{author}{Schinko, T.} \emph{et~al.}
\newblock \bibinfo{journal}{\bibinfo{title}{Economy-wide effects of coastal flooding due to sea level rise: a multi-model simultaneous treatment of mitigation, adaptation, and residual impacts}}.
\newblock {\emph{\JournalTitle{Environmental Research Communications}}} \textbf{\bibinfo{volume}{2}}, \bibinfo{pages}{015002} (\bibinfo{year}{2020}).

\bibitem{haasnoot2021pathways}
\bibinfo{author}{Haasnoot, M.}, \bibinfo{author}{Lawrence, J.} \& \bibinfo{author}{Magnan, A.~K.}
\newblock \bibinfo{journal}{\bibinfo{title}{Pathways to coastal retreat}}.
\newblock {\emph{\JournalTitle{Science}}} \textbf{\bibinfo{volume}{372}}, \bibinfo{pages}{1287--1290} (\bibinfo{year}{2021}).

\bibitem{kirezci2020projections}
\bibinfo{author}{Kirezci, E.} \emph{et~al.}
\newblock \bibinfo{journal}{\bibinfo{title}{Projections of global-scale extreme sea levels and resulting episodic coastal flooding over the 21st century}}.
\newblock {\emph{\JournalTitle{Scientific reports}}} \textbf{\bibinfo{volume}{10}}, \bibinfo{pages}{11629} (\bibinfo{year}{2020}).

\bibitem{hinkel2014}
\bibinfo{author}{Hinkel, J.} \emph{et~al.}
\newblock \bibinfo{journal}{\bibinfo{title}{Coastal flood damage and adaptation costs under 21st century sea-level rise}}.
\newblock {\emph{\JournalTitle{Proceedings of the National Academy of Sciences}}} \textbf{\bibinfo{volume}{111}}, \bibinfo{pages}{3292--3297} (\bibinfo{year}{2014}).

\bibitem{merel2021climate}
\bibinfo{author}{M{\'e}rel, P.} \& \bibinfo{author}{Gammans, M.}
\newblock \bibinfo{journal}{\bibinfo{title}{Climate econometrics: Can the panel approach account for long-run adaptation?}}
\newblock {\emph{\JournalTitle{American Journal of Agricultural Economics}}} \textbf{\bibinfo{volume}{103}}, \bibinfo{pages}{1207--1238} (\bibinfo{year}{2021}).

\bibitem{auffhammer2018quantifying}
\bibinfo{author}{Auffhammer, M.}
\newblock \bibinfo{journal}{\bibinfo{title}{Quantifying economic damages from climate change}}.
\newblock {\emph{\JournalTitle{Journal of Economic Perspectives}}} \textbf{\bibinfo{volume}{32}}, \bibinfo{pages}{33--52} (\bibinfo{year}{2018}).

\bibitem{burke2015global}
\bibinfo{author}{Burke, M.}, \bibinfo{author}{Hsiang, S.~M.} \& \bibinfo{author}{Miguel, E.}
\newblock \bibinfo{journal}{\bibinfo{title}{Global non-linear effect of temperature on economic production}}.
\newblock {\emph{\JournalTitle{Nature}}} \textbf{\bibinfo{volume}{527}}, \bibinfo{pages}{235--239} (\bibinfo{year}{2015}).

\bibitem{novavckova2018effects}
\bibinfo{author}{Nov{\'a}{\v{c}}kov{\'a}, M.} \& \bibinfo{author}{Tol, R.~S.}
\newblock \bibinfo{journal}{\bibinfo{title}{Effects of sea level rise on economy of the united states}}.
\newblock {\emph{\JournalTitle{Journal of Environmental Economics and Policy}}} \textbf{\bibinfo{volume}{7}}, \bibinfo{pages}{85--115} (\bibinfo{year}{2018}).

\bibitem{RN10}
\bibinfo{author}{Chatzivasileiadis, T.}, \bibinfo{author}{Estrada, F.}, \bibinfo{author}{Hofkes, M.} \& \bibinfo{author}{Tol, R.}
\newblock \bibinfo{journal}{\bibinfo{title}{Systematic sensitivity analysis of the full economic impacts of sea level rise}}.
\newblock {\emph{\JournalTitle{Computational Economics}}} \textbf{\bibinfo{volume}{53}}, \bibinfo{pages}{1183--1217} (\bibinfo{year}{2019}).

\bibitem{RN9}
\bibinfo{author}{Pycroft, J.}, \bibinfo{author}{Abrell, J.} \& \bibinfo{author}{Ciscar, J.-C.}
\newblock \bibinfo{journal}{\bibinfo{title}{The global impacts of extreme sea-level rise: A comprehensive economic assessment}}.
\newblock {\emph{\JournalTitle{Environmental and Resource Economics}}} \bibinfo{pages}{1--29} (\bibinfo{year}{2015}).

\bibitem{RN8}
\bibinfo{author}{Eboli, F.}, \bibinfo{author}{Parrado, R.} \& \bibinfo{author}{Roson, R.}
\newblock \bibinfo{journal}{\bibinfo{title}{Climate-change feedback on economic growth: explorations with a dynamic general equilibrium model}}.
\newblock {\emph{\JournalTitle{Environment and Development Economics}}} \textbf{\bibinfo{volume}{15}}, \bibinfo{pages}{515--533} (\bibinfo{year}{2010}).

\bibitem{RN4}
\bibinfo{author}{Bosello, F.}, \bibinfo{author}{Roson, R.} \& \bibinfo{author}{Tol, R.~S.}
\newblock \bibinfo{journal}{\bibinfo{title}{Economy-wide estimates of the implications of climate change: Sea level rise}}.
\newblock {\emph{\JournalTitle{Environmental and Resource Economics}}} \textbf{\bibinfo{volume}{37}}, \bibinfo{pages}{549--571} (\bibinfo{year}{2007}).

\bibitem{RN7}
\bibinfo{author}{Bigano, A.}, \bibinfo{author}{Bosello, F.}, \bibinfo{author}{Roson, R.} \& \bibinfo{author}{Tol, R.~S.}
\newblock \bibinfo{journal}{\bibinfo{title}{Economy-wide impacts of climate change: A joint analysis for sea level rise and tourism}}.
\newblock {\emph{\JournalTitle{Mitigation and Adaptation Strategies for Global Change}}} \textbf{\bibinfo{volume}{13}}, \bibinfo{pages}{765--791} (\bibinfo{year}{2008}).

\bibitem{melet2021european}
\bibinfo{author}{Melet, A.} \emph{et~al.}
\newblock \bibinfo{journal}{\bibinfo{title}{European copernicus services to inform on sea-level rise adaptation: current status and perspectives}}.
\newblock {\emph{\JournalTitle{Frontiers in Marine Science}}} \textbf{\bibinfo{volume}{8}}, \bibinfo{pages}{703425} (\bibinfo{year}{2021}).

\bibitem{mcevoy2021european}
\bibinfo{author}{McEvoy, S.}, \bibinfo{author}{Haasnoot, M.} \& \bibinfo{author}{Biesbroek, R.}
\newblock \bibinfo{journal}{\bibinfo{title}{How are european countries planning for sea level rise?}}
\newblock {\emph{\JournalTitle{Ocean \& Coastal Management}}} \textbf{\bibinfo{volume}{203}}, \bibinfo{pages}{105512} (\bibinfo{year}{2021}).

\bibitem{Fankhauser2005}
\bibinfo{author}{Fankhauser, S.} \& \bibinfo{author}{Tol, R. S.~J.}
\newblock \bibinfo{journal}{\bibinfo{title}{On climate change and economic growth}}.
\newblock {\emph{\JournalTitle{Resource and Energy Economics}}} \textbf{\bibinfo{volume}{27}}, \bibinfo{pages}{1--17} (\bibinfo{year}{2005}).

\bibitem{magnan2023status}
\bibinfo{author}{Magnan, A.~K.} \emph{et~al.}
\newblock \bibinfo{journal}{\bibinfo{title}{Status of global coastal adaptation}}.
\newblock {\emph{\JournalTitle{Nature Climate Change}}} \bibinfo{pages}{1--9} (\bibinfo{year}{2023}).

\bibitem{dell2009temperature}
\bibinfo{author}{Dell, M.}, \bibinfo{author}{Jones, B.~F.} \& \bibinfo{author}{Olken, B.~A.}
\newblock \bibinfo{journal}{\bibinfo{title}{Temperature and income: reconciling new cross-sectional and panel estimates}}.
\newblock {\emph{\JournalTitle{American Economic Review}}} \textbf{\bibinfo{volume}{99}}, \bibinfo{pages}{198--204} (\bibinfo{year}{2009}).

\bibitem{PSMSL_2023}
\bibinfo{author}{for Mean Sea Level~(PSMSL), P.~S.}
\newblock \bibinfo{title}{Tide gauge data}.
\newblock \bibinfo{howpublished}{\url{http://www.psmsl.org/data/obtaining/}} (\bibinfo{year}{2023}).
\newblock \bibinfo{note}{Accessed: 2023-07-24}.

\bibitem{holgate2013new}
\bibinfo{author}{Holgate, S.~J.} \emph{et~al.}
\newblock \bibinfo{journal}{\bibinfo{title}{New data systems and products at the permanent service for mean sea level}}.
\newblock {\emph{\JournalTitle{Journal of Coastal Research}}} \textbf{\bibinfo{volume}{29}}, \bibinfo{pages}{493--504} (\bibinfo{year}{2013}).

\bibitem{roses2021regional}
\bibinfo{author}{Ros{\'e}s, J.~R.} \& \bibinfo{author}{Wolf, N.}
\newblock \bibinfo{journal}{\bibinfo{title}{Regional growth and inequality in the long-run: Europe, 1900--2015}}.
\newblock {\emph{\JournalTitle{Oxford Review of Economic Policy}}} \textbf{\bibinfo{volume}{37}}, \bibinfo{pages}{17--48} (\bibinfo{year}{2021}).

\bibitem{dell2008climate}
\bibinfo{author}{Dell, M.}, \bibinfo{author}{Jones, B.~F.} \& \bibinfo{author}{Olken, B.~A.}
\newblock \bibinfo{title}{Climate change and economic growth: Evidence from the last half century}.
\newblock \bibinfo{type}{Tech. Rep.}, \bibinfo{institution}{National Bureau of Economic Research} (\bibinfo{year}{2008}).

\bibitem{psmsl_2017}
\bibinfo{author}{{Permanent Service for Mean Sea Level (PSMSL)}}.
\newblock \bibinfo{title}{Permanent service for mean sea level: Further information}.
\newblock \bibinfo{howpublished}{\url{https://psmsl.org/data/obtaining/}} (\bibinfo{year}{2017}).
\newblock \bibinfo{note}{Accessed: 2023-11-07}.

\bibitem{lincked2}
\bibinfo{author}{Lincke, D.} \emph{et~al.}
\newblock \bibinfo{title}{D2. 3 impacts on infrastructure, built environment, and transport}.
\newblock \bibinfo{howpublished}{\url{https://doi.org/10.5281/zenodo.5703656}}.

\bibitem{hinkel2014coastal}
\bibinfo{author}{Hinkel, J.} \emph{et~al.}
\newblock \bibinfo{journal}{\bibinfo{title}{Coastal flood damage and adaptation costs under 21st century sea-level rise}}.
\newblock {\emph{\JournalTitle{Proceedings of the National Academy of Sciences}}} \textbf{\bibinfo{volume}{111}}, \bibinfo{pages}{3292--3297} (\bibinfo{year}{2014}).

\bibitem{lincke2021coastal}
\bibinfo{author}{Lincke, D.} \& \bibinfo{author}{Hinkel, J.}
\newblock \bibinfo{journal}{\bibinfo{title}{Coastal migration due to 21st century sea-level rise}}.
\newblock {\emph{\JournalTitle{Earth's Future}}} \textbf{\bibinfo{volume}{9}}, \bibinfo{pages}{e2020EF001965} (\bibinfo{year}{2021}).

\bibitem{oecd2019responding}
\bibinfo{author}{OECD}.
\newblock \bibinfo{title}{Responding to rising seas oecd country approaches to tackling coastal risks} (\bibinfo{year}{2019}).

\bibitem{berrang2021systematic}
\bibinfo{author}{Berrang-Ford, L.} \emph{et~al.}
\newblock \bibinfo{journal}{\bibinfo{title}{A systematic global stocktake of evidence on human adaptation to climate change}}.
\newblock {\emph{\JournalTitle{Nature Climate Change}}} \textbf{\bibinfo{volume}{11}}, \bibinfo{pages}{989--1000} (\bibinfo{year}{2021}).

\bibitem{Hamilton2007}
\bibinfo{author}{Hamilton, J.~M.}
\newblock \bibinfo{journal}{\bibinfo{title}{Coastal landscape and the hedonic price of accommodation}}.
\newblock {\emph{\JournalTitle{Ecological Economics}}} \textbf{\bibinfo{volume}{62}}, \bibinfo{pages}{594--602}, \doiprefix\url{https://doi.org/10.1016/j.ecolecon.2006.08.001} (\bibinfo{year}{2007}).

\bibitem{abdelhafez2022hidden}
\bibinfo{author}{Abdelhafez, M.~A.}, \bibinfo{author}{Ellingwood, B.} \& \bibinfo{author}{Mahmoud, H.}
\newblock \bibinfo{journal}{\bibinfo{title}{Hidden costs to building foundations due to sea level rise in a changing climate}}.
\newblock {\emph{\JournalTitle{Scientific reports}}} \textbf{\bibinfo{volume}{12}}, \bibinfo{pages}{14020} (\bibinfo{year}{2022}).

\bibitem{Yohe2002}
\bibinfo{author}{Yohe, G.~W.} \& \bibinfo{author}{Tol, R. S.~J.}
\newblock \bibinfo{journal}{\bibinfo{title}{Indicators for social and economic coping capacity\textemdash moving towards a working definition of adaptive capacity}}.
\newblock {\emph{\JournalTitle{Global Environmental Change}}} \textbf{\bibinfo{volume}{12}}, \bibinfo{pages}{25--40} (\bibinfo{year}{2002}).

\bibitem{stiglitz2009report}
\bibinfo{author}{Stiglitz, J.~E.}, \bibinfo{author}{Sen, A.}, \bibinfo{author}{Fitoussi, J.-P.} \emph{et~al.}
\newblock \bibinfo{title}{Report by the commission on the measurement of economic performance and social progress} (\bibinfo{year}{2009}).

\bibitem{chancel2014beyond}
\bibinfo{author}{Chancel, L.}, \bibinfo{author}{Thiry, G.} \& \bibinfo{author}{Demailly, D.}
\newblock \bibinfo{journal}{\bibinfo{title}{Beyond-gdp indicators: to what end}}.
\newblock {\emph{\JournalTitle{Lessons learnt from six national experiences. Study}}} \bibinfo{pages}{30} (\bibinfo{year}{2014}).

\bibitem{EuropeanCommission2021}
\bibinfo{author}{Terzi, A.}
\newblock \bibinfo{title}{Economic policy-making beyond gdp: An introduction}.
\newblock \bibinfo{type}{Discussion Paper} \bibinfo{number}{142}, \bibinfo{institution}{European Commission} (\bibinfo{year}{2021}).
\newblock \bibinfo{note}{Accessed: 2023-11-07}.

\bibitem{roses2018economic}
\bibinfo{author}{Ros{\'e}s, J.~R.} \& \bibinfo{author}{Wolf, N.}
\newblock \bibinfo{title}{The economic development of europe's regions: A quantitative history since 1900} (\bibinfo{year}{2018}).

\bibitem{wolff2023setback}
\bibinfo{author}{Wolff, C.}, \bibinfo{author}{Bonatz, H.} \& \bibinfo{author}{Vafeidis, A.~T.}
\newblock \bibinfo{journal}{\bibinfo{title}{Setback zones can effectively reduce exposure to sea-level rise in europe}}.
\newblock {\emph{\JournalTitle{Scientific Reports}}} \textbf{\bibinfo{volume}{13}}, \bibinfo{pages}{5515} (\bibinfo{year}{2023}).

\bibitem{mohaddes2023climate}
\bibinfo{author}{Mohaddes, K.}, \bibinfo{author}{Ng, R.~N.}, \bibinfo{author}{Pesaran, M.~H.}, \bibinfo{author}{Raissi, M.} \& \bibinfo{author}{Yang, J.-C.}
\newblock \bibinfo{journal}{\bibinfo{title}{Climate change and economic activity: evidence from us states}}.
\newblock {\emph{\JournalTitle{Oxford Open Economics}}} \textbf{\bibinfo{volume}{2}} (\bibinfo{year}{2023}).

\bibitem{Eurostat_2023a}
\bibinfo{author}{Eurostat}.
\newblock \bibinfo{title}{Population on 1 january by age, sex and nuts 2 region}.
\newblock \bibinfo{howpublished}{\url{https://ec.europa.eu/eurostat/databrowser/view/DEMO_R_D2JAN/default/table?lang=en}} (\bibinfo{year}{2023}).
\newblock \bibinfo{note}{Accessed: 2023-07-29}.

\bibitem{Eurostat_GDP_2023}
\bibinfo{author}{Eurostat}.
\newblock \bibinfo{title}{Gross domestic product (gdp) at current market prices by nuts 2 regions}.
\newblock \bibinfo{howpublished}{\url{https://ec.europa.eu/eurostat/databrowser/view/NAMA_10R_2GDP/default/table?lang=en}} (\bibinfo{year}{2023}).
\newblock \bibinfo{note}{Accessed: 2023-07-29}.

\bibitem{vafeidis2008new}
\bibinfo{author}{Vafeidis, A.~T.} \emph{et~al.}
\newblock \bibinfo{journal}{\bibinfo{title}{A new global coastal database for impact and vulnerability analysis to sea-level rise}}.
\newblock {\emph{\JournalTitle{Journal of coastal research}}} \textbf{\bibinfo{volume}{24}}, \bibinfo{pages}{917--924} (\bibinfo{year}{2008}).

\bibitem{peltier2004global}
\bibinfo{author}{Peltier, W.~R.}
\newblock \bibinfo{journal}{\bibinfo{title}{Global glacial isostasy and the surface of the ice-age earth: the ice-5g (vm2) model and grace}}.
\newblock {\emph{\JournalTitle{Annu. Rev. Earth Planet. Sci.}}} \textbf{\bibinfo{volume}{32}}, \bibinfo{pages}{111--149} (\bibinfo{year}{2004}).

\bibitem{nicholls2021global}
\bibinfo{author}{Nicholls, R.~J.} \emph{et~al.}
\newblock \bibinfo{journal}{\bibinfo{title}{A global analysis of subsidence, relative sea-level change and coastal flood exposure}}.
\newblock {\emph{\JournalTitle{Nature Climate Change}}} \textbf{\bibinfo{volume}{11}}, \bibinfo{pages}{338--342} (\bibinfo{year}{2021}).

\bibitem{bachner2022macroeconomic}
\bibinfo{author}{Bachner, G.}, \bibinfo{author}{Lincke, D.} \& \bibinfo{author}{Hinkel, J.}
\newblock \bibinfo{journal}{\bibinfo{title}{The macroeconomic effects of adapting to high-end sea-level rise via protection and migration}}.
\newblock {\emph{\JournalTitle{Nature Communications}}} \textbf{\bibinfo{volume}{13}}, \bibinfo{pages}{5705} (\bibinfo{year}{2022}).

\bibitem{o2017roads}
\bibinfo{author}{O’Neill, B.~C.} \emph{et~al.}
\newblock \bibinfo{journal}{\bibinfo{title}{The roads ahead: Narratives for shared socioeconomic pathways describing world futures in the 21st century}}.
\newblock {\emph{\JournalTitle{Global environmental change}}} \textbf{\bibinfo{volume}{42}}, \bibinfo{pages}{169--180} (\bibinfo{year}{2017}).

\bibitem{IPCC_2021}
\bibinfo{author}{[IPCC]}.
\newblock \bibinfo{title}{Climate change 2021: The physical science basis. contribution of working group i to the sixth assessment report of the intergovernmental panel on climate change}, \doiprefix\url{10.1017/9781009157896} (\bibinfo{year}{2021}).

\end{thebibliography}

\section*{Acknowledgements}

\section*{Author contributions statement}
TC collected the data, designed the empirical strategy, ran the estimation, and wrote the first draft. RT advised on the empirical strategy and edited the manuscript. ICA revised the manuscript, JH and DL provided the data.

\section*{Competing interests}
The authors declare no competing interests.

\section*{Data availability}
The datasets generated during and/or analysed during the current study will be available in the Zenodo repository.

\end{document}